\documentclass[journal=jctcce,manuscript=article]{achemso}

\usepackage{chemformula} 
\usepackage[T1]{fontenc} 
\usepackage{booktabs}
\usepackage{hyperref}
\usepackage{orcidlink}
\usepackage{svg}
\usepackage[dvipsnames]{xcolor}
\usepackage{comment}

\usepackage[normalem]{ulem}
\usepackage{xspace}

\usepackage{multirow}
\author{Surender Kumar\orcidlink{0009-0000-3072-5633}}
\affiliation{Institute for Condensed Matter Physics and Optics, Friedrich-Schiller-Universit\"at Jena, 07743 Jena, Germany}
\email{surendermohinder@gmail.com}

\author{Martin Th\"ummler\orcidlink{0000-0001-8171-7895}}
\affiliation{%
	Institute of Physical Chemistry, Friedrich-Schiller-Universit\"at Jena, 07743 Jena, Germany
}%

\author{Alexander Croy\orcidlink{0000-0001-9296-9350}}
\affiliation{%
	Institute of Physical Chemistry, Friedrich-Schiller-Universit\"at Jena, 07743 Jena, Germany
}%
\email{alexander.croy@uni-jena.de}
\author{Stefanie Gräfe\orcidlink{0000-0002-1747-5809}}
\affiliation{%
	Institute of Physical Chemistry, Friedrich-Schiller-Universit\"at Jena, 07743 Jena, Germany
}
\alsoaffiliation{Abbe Center of Photonics, Friedrich-Schiller-Universit\"at Jena, 07745, Jena, Germany}
\altaffiliation{Fraunhofer Institute of Applied Optics and Precision Engineering (Fraunhofer IOF), Albert-Einstein-Str. 7, 07745 Jena, Germany}%
\author{Caterina Cocchi\orcidlink{0000-0002-9243-9461}}
\affiliation{Institute for Condensed Matter Physics and Optics, Friedrich-Schiller-Universit\"at Jena, 07743 Jena, Germany}
\alsoaffiliation{Abbe Center of Photonics, Friedrich-Schiller-Universit\"at Jena, 07745, Jena, Germany}
\email{caterina.cocchi@uni-jena.de}

\title{Multiscale Quasiparticle Electronic Structure and Excitonic Properties of CdSe Nanoclusters}

\keywords{CdSe, Nanoclusters, GW BSE, Exciton, Optical properties}

\begin{document}

\begin{tocentry}
\centering
\includegraphics[]{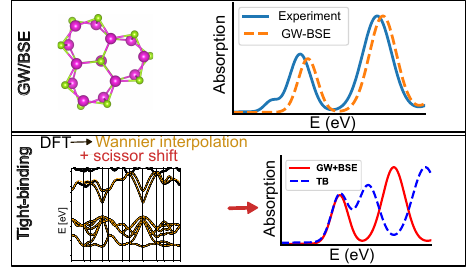}
\end{tocentry}

\newpage
\begin{abstract}
Quantum confinement in stoichiometric $\text{Cd}_n\text{Se}_n$ nanoclusters dramatically attenuates electronic screening, driving a delicate, size-dependent competition between quasiparticle self-energy corrections ($\Delta_{\mathrm{QP}}$) and exciton binding energies ($E_b$). Here, we present a $GW$/BSE study across a representative size series ($n = 3, 6, 13, 33$) and leverage it to validate a scalable atomistic tight-binding (TB) framework derived from first principles. Our results demonstrate that 1--2~eV spectral blueshifts previously reported in the literature arise from single-particle $GW$ convergence artifacts rather than deficiencies in the electron–hole kernels. We show that the near-perfect cancellation between $\Delta_{\mathrm{QP}}$ and $E_b$ breaks down as cluster volume increases, driven by the rapid onset of dielectric screening attenuating $E_b$ faster than $\Delta_{\mathrm{QP}}$ and leading to a pronounced divergence from mean-field predictions. Spatial inverse participation ratio analysis of the electronic structure reveals that optical suppression of fundamental pre-peaks stems from a severe spatial mismatch between localized valence orbitals and delocalized conduction states. Finally, we demonstrate that the confinement-induced scaling of the quasiparticle gap and the optical onset is accurately reproduced by a scissor-corrected, DFT-parameterized TB model. As such, this work provides a quantitative multiscale roadmap for embedding effective many-body effects kernels into computationally efficient models, enabling reliable optical predictions for realistic semiconducting nanostructures containing up to thousands of atoms.
 
\end{abstract}

\newpage
\section{INTRODUCTION}

The continuous efforts to exploit linear\cite{Talapin2010,ibanez2025,Arquer2021} and nonlinear optical responses\cite{Hanamura1998,Banfi1998,nakagawa2022,krishna2023,Ganeev2022} in semiconducting nanostructures, with dimensions below 10 nanometers (nm), have established them as a premier platform for studying light–matter interactions at the nanoscale. While their linear properties stimulate applications in photovoltaics\cite{lee2008} and light-emitting technologies\cite{panfil2018,Jang2010}, optical nonlinearities open opportunities for ultrafast and quantum photonic devices\cite{Petruska2003,Srinivasan2007,Loo2012,Javadi2015,Wu2025}. Thanks to their extensive experimental characterization driven by enhanced light-matter coupling~\cite{Kagan2016,Kagan2021, nakagawa2022,krishna2023,Ganeev2022,jacobsohn2000,Seo2006,majdabadi2015}, cadmium selenide (CdSe) nanocrystals have been regarded with special interest in this field. At sub-nanometer sizes, the almost complete collapse of bulk-like dielectric screening boosts the interplay between quantum confinement and many-body interactions, making optical responses exceptionally sensitive to edge effects and the localized character of the frontier Se $4p$ and Cd $5s$ states\cite{Puzder2004,mcisaac2023,kilina2016,steenbock2024,houtepen2017}. These entangled characteristics render CdSe nanocrystals promising building blocks for next-generation optoelectronic, photonic, and quantum technologies\cite{Aichele2003,Prabhakaran2012,Grim2015,lin2017,Efros2021}.

Accurately describing excited states in semiconducting nanoclusters  remains, however, a formidable theoretical hurdle. While wave-function-based quantum chemistry methods are the gold standard for accessing excitation spectra of small molecules and quantum dots\cite{Bartlett2007,Sneskov2012,sarkar2021}, their computational cost becomes exorbitant when the radius expands beyond a few Angstroms\cite{werner2011}. Density functional theory (DFT)\cite{Hohenberg1964,Kohn1965} and its time-dependent extension (TDDFT)\cite{Runge1984}, as well as accurate tight-binding (TB) implementations~\cite{Peschel2022,thummler_modeling_2026}, offer computationally viable alternatives for handling nanostructures atomistically. However, standard exchange–correlation functionals provide only an approximate treatment of electron correlations \cite{Burke2012} and often fail to adequately describe the interplay between quasiparticle self-energy corrections (many-body effects on electron addition and removal energies) and excitonic interactions (electron–hole Coulomb interactions associated with electronic excitations), particularly under strong quantum confinement~\cite{Jose2006,Fischer2012,Cui2015,Bhati2022}.This limitation often leads to sizable inconsistencies with respect to optical measurements~\cite{Del2011,Jose2006}. 

Many-body perturbation theory (MBPT)\cite{Onida2002}, combining the \textit{GW} approximation\cite{Hedin1965} for the electronic self-energy with the solution of the Bethe–Salpeter equation (BSE)\cite{BSE1951} for the correlated electron–hole excitations, has emerged as a valuable alternative\cite{Tiago2006,Cocchi2015,jacquemin2015} , particularly in light of recent developments extending a formalism traditionally established for periodic solids to zero-dimensional systems\cite{Rocca2010,Baumeier2012,Hirose2015,Bruneval2015}. Beyond reliably predicting the optical properties of organic molecules\cite{Palummo2009} and inorganic nanostructures~\cite{Govoni2015,lopez2008}, MBPT yields direct access to the microscopic nature of the electronic transitions, including their spatial delocalization and single-particle orbital (i.e. atomic orbitals) composition. Such insights are of paramount importance for achieving a fundamental understanding of the light-matter interactions in nanoconfined semiconductors.

CdSe nanoclusters were among the first non-periodic (or confined) systems investigated with the {\textit{GW}/BSE} formalism~\cite{Del2006,lopez2008,noguchi2012}. Those early studies reported systematic blue-shifts $\geq$~1~eV of excitation energies compared to experiments~\cite{lopez2008,noguchi2012}. These historical deviations reflect both the evolution of computing hardware and software, progressively enabling increasing numerical convergence and more sophisticated approximations\cite{Govoni2015,wilhelm2018,del2019,umari2022,gao2024}, and the inherent challenge of capturing many-body interactions under quantum confinement. A comprehensive understanding of these systems therefore requires the combined accuracy of first-principles many-body theory with scalability across increasing sizes. Atomistic TB models fulfill this role by retaining an explicit orbital basis to describe quantum confinement trends and electronic structures, provided they are systematically benchmarked against rigorous quasiparticle calculations. When appropriately parameterized, they offer a practical foundation for many-body descriptions of linear excitations of semiconductor nanostructures of realistic sizes otherwise inaccessible for fully \textit{ab initio} workflows, with feasible extensions to the nonlinear regime.

In this work, we investigate electronic and optical excitations of four stoichiometric $\text{Cd}_n\text{Se}_n$ nanoclusters ($n = 3$, 6, 13, and 33), including a 2-fold coordinated six-membered ring ($\text{Cd}_{3}\text{Se}_{3}$), a stacked 3-fold coordinated prism ($\text{Cd}_{6}\text{Se}_{6}$), and the larger wurtzite nanocrystals ($\text{Cd}_{13}\text{Se}_{13}$ and $\text{Cd}_{33}\text{Se}_{33}$). Using the \textit{GW}/BSE formalism, we determine quasiparticle and excitonic (optical) spectra, benchmarking our results against available theoretical and experimental references. We leverage these findings to evaluate how a  scissor-corrected (rigidly shifted electronic energy levels), DFT-parameterized atomistic TB model reproduces electronic and optical properties at increasing cluster size. We show that while the  quasiparticle gap (fundamental electronic gap) and optical absorption onset are captured remarkably well even at the independent-particle (mean-field) level  currently available in the adopted TB framework, an explicit treatment of electron--hole correlations remains crucial for properly describing higher-energy excitations. By establishing the accuracy of the \textit{GW}/BSE approach for CdSe nanoclusters and validating a complementary atomistic framework, this work offers a practical and predictive route for modeling optical properties in quantum-confined semiconducting nanostructures.

\section{COMPUTATIONAL DETAILS}
Ground-state properties were determined using density-functional theory (DFT)\cite{Hohenberg1964,Kohn1965} as implemented in the \texttt{Quantum ESPRESSO} package~\cite{giannozzi2017}. Core-electron interactions were described using optimized norm-conserving Vanderbilt pseudopotentials~\cite{Hamann2013,Setten2018} without including spin–orbit coupling (SOC), while exchange-correlation effects were modeled with the Perdew-Burke-Ernzerhof (PBE) functional~\cite{Perdew1996}. To simulate non-periodic nanoclusters and remove spurious electrostatic interactions between replicas, we modeled the systems in a cubic box with a minimum vacuum layer of at least 10~\AA{} in all directions and sampled the Brillouin zone at the $\Gamma$-point only. Structural relaxations were carried out with convergence criteria for total energies and interatomic forces of $10^{-6}$~Ry and $10^{-3}$~Ry/Bohr, respectively. A kinetic energy cutoff of 52~Ry for the plane-wave basis set was found sufficient to converge the Kohn-Sham gap to within 10~meV. 

To analyze state localization across the energy spectrum, the DFT electronic structure was post-processed in terms of the inverse participation ratio (IPR) for a given state $n$, defined as~\cite{Murphy2011,Calixto2015,Kumar2025}
\begin{equation}
    \mathrm{IPR}_{n} = \frac{\sum_{a} |c_{na}|^4}{\left(\sum_a |c_{na}|^2\right)^2},
    \label{eq:ipr}
\end{equation}
where $a$ represents the atomic site index and $c_{na}$ denotes the corresponding wave-function projection coefficients. Large IPRs indicate pronounced state localization, whereas smaller values are associated with delocalized states~\cite{Kumar2026}.

Many-body perturbation theory calculations were subsequently performed using the \texttt{WEST} code~\cite{Govoni2015}. The static dielectric screening was computed using the projective dielectric eigendecomposition (PDEP) technique\cite{Wilson2008}, which avoids explicit summations over empty states by representing the polarizability operator in a reduced eigenbasis~\cite{Govoni2015}. 
Single-shot $G_0W_0$ calculations~\cite{Hybertsen1986} were converged with an accuracy of approximately 10~meV by including a number of PDEP eigenvectors equal to ten times the number of occupied states (convergence tests in Figure~S1). To accelerate the evaluation of nonlocal Coulomb integrals, the occupied Kohn–Sham orbitals were transformed into maximally localized Wannier functions, applying an overlap threshold of $10^{-3}$~\cite{Monti2025}. 
Optical excitations were subsequently computed by solving the BSE~\cite{BSE1951} within the Tamm-Dancoff approximation~\cite{HIRATA1999}. The lowest 250 eigenvalues were obtained from the Liouville superoperator using the Davidson algorithm, bypassing the explicit construction and diagonalization of a Bethe–Salpeter Hamiltonian matrix in a transition basis~\cite{Rocca2010}. SOC was not included as it is expected to induce merely a minor, rigid energy shift in the optical onset without affecting the physical trends or the magnitude of the exciton binding energies~\cite{Bui2020,Steenbock2023}.

To complement the first-principles many-body framework for accessing nanocrystals with a larger size, an atomistic TB model was parametrized from a bulk DFT calculation for wurtzite CdSe performed using \texttt{Quantum ESPRESSO}~\cite{giannozzi2017} with the PBE functional~\cite{Perdew1996} and norm-conserving pseudopotentials~\cite{Setten2018,hamann_optimized_2013}. The geometry and ground-state density were optimized on a $10\times10 \times 10$ grid adopting a plane-wave cut-off of $150\,\mathrm{Ry}$. Real-space Hamiltonian matrix elements for an 8-band model (comprising six valence and two conduction bands per unit cell) were generated using \texttt{Wannier90}~\cite{pizzi_wannier90_2020} on a $9\times 9\times 9$ $k$-grid with dipole matrix elements derived according to Ref.~\citenum{thummler2026self}. 
A rigid scissor operator was applied to align the bulk band gap with the experimental value of $1.76\,\mathrm{eV}$ \cite{ninomiya_optical_1995}.
 The considered geometries for the TB model are set up from the periodic structure cut at unit cell boundaries, corresponding by construction to the bulk limit. Further details are reported in Ref.~\citenum{thummler_modeling_2026} and in the Supporting Information (Section S2).
Size-dependent optical absorption spectra were subsequently calculated by propagating the density matrix under a linearly polarized electric field (see further details in the Supporting Information).

\section{RESULTS AND DISCUSSION}
\begin{figure}
    \centering
    \includegraphics[width=0.75\linewidth]{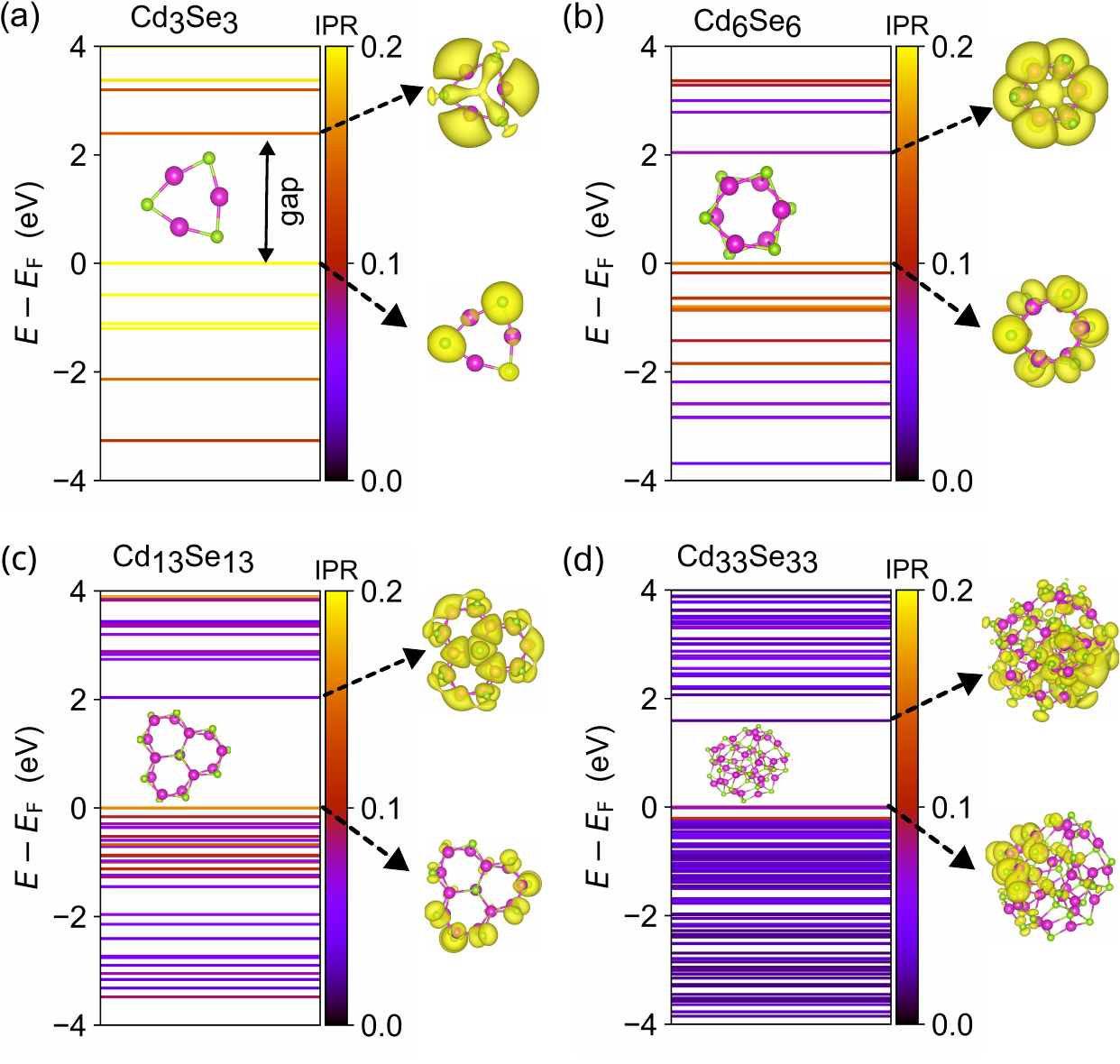}
   \caption{Single-particle energy levels colored according to their inverse participation ratio (IPR) computed for (a) Cd$_3$Se$_3$, (b) Cd$_6$Se$_6$, (c) Cd$_{13}$Se$_{13}$, and (d) Cd$_{33}$Se$_{33}$. The squared wavefunctions of the HOMO and LUMO in the optimized cluster geometries (see insets) are visualized next to each graph. The energy levels are offset to the Fermi energy (E$_F$), aligned to the HOMO level.}
    \label{fig:stru_level}
\end{figure}

The relaxed geometries of the four stoichiometric $\text{Cd}_n\text{Se}_n$ nanoclusters ($n = 3$, 6, 13, and 33) indicate a clear structural evolution from molecular rings to crystalline environments (see insets of Figure~\ref{fig:stru_level}).  The smallest system, $\text{Cd}_{3}\text{Se}_{3}$, forms a planar six-membered ring with 2-fold coordinated atoms, while $\text{Cd}_{6}\text{Se}_{6}$ adopts a stacked prism geometry in which all atoms are 3-fold coordinated. The larger $\text{Cd}_{13}\text{Se}_{13}$ and $\text{Cd}_{33}\text{Se}_{33}$ nanoclusters display core wurtzite configurations\cite{Kumar2024}, where internal bulk-like atoms maintain 4-fold coordination, whereas unpassivated surface atoms exhibit 3-fold coordination. We note in passing that we chose unpassivated clusters on purpose to maintain consistency with earlier GW/BSE studies~\cite{noguchi2012}. Upon structural optimization, these unpassivated surfaces undergo significant structural reconstruction, a well-known characteristic of II--VI semiconductor clusters~\cite{Puzder2004,Fischer2012,Bhati2022}.

The absence of surface ligands and the accompanying variations in coordination environments directly shape the single-particle electronic landscapes (Figure~\ref{fig:stru_level}).
As the cluster size increases, the band gap decreases, reflecting well-known quantum-confinement considerations and established experimental trends~\cite{kasuya2004,Jose2006}. In all cases, the lowest unoccupied molecular orbital (LUMO) is singly degenerate, while the highest occupied molecular orbital (HOMO) is nearly degenerate~\cite{rovasio2026}. The minor energy splittings between the HOMO and HOMO-1 levels ($\sim$2~meV, reaching 15~meV for Cd${_{33}}$Se${_{33}}$) arise directly from symmetry-breaking surface reconstructions. 

The HOMO manifold is primarily composed of Se 4$p$ orbitals, whereas the LUMO originates mainly from Cd 5$s$ states, following the composition of II–VI semiconductors~\cite{fenoll2023,rovasio2026}. Across increasing cluster size, the energy levels become more closely spaced, eventually evolving toward a continuum of bands with further increase in size. The energetic splitting between the LUMO and LUMO+1 is generally larger than between the HOMO and the HOMO–1 or HOMO–2. As confirmed by our IPR analysis (Figure~\ref{fig:stru_level}), in all studied systems, the LUMO is relatively delocalized across the entire cluster, whereas the HOMO being confined near the surface, remains significantly more localized than the LUMO. Occupied frontier states are significantly more localized than deeper valence states and the lowest unoccupied orbitals. This localization is associated with surface trap formation, as extensively discussed in previous theoretical and experimental studies~\cite{kilina2009,Bhati2022,rovasio2026,Bieniek2025,goldzak2021,mcisaac2023}.

The self-energy contribution significantly increases the band-gap size (Table~\ref{tab:gaps_binding}). The quasiparticle (QP) correction, defined as $\Delta_{\mathrm{QP}} = E_g^{GW}-E_g^{DFT}$, shows a strong size dependence, which decreases monotonically from 4.21~eV in Cd${_3}$Se${_3}$ to 2.49~eV in Cd${_{33}}$Se${_{33}}$. This trend highlights the extreme sensitivity of many-body correlations to real-space confinement: as the cluster volume shrinks, the almost complete collapse of dielectric screening enhances the electron-electron repulsive self-energy, which gradually decreases as the system size scales up, asymptotically approaching the bulk-like screening boundaries~\cite{Cocchi2015}. 

For the smallest clusters, $\text{Cd}_3\text{Se}_3$ and $\text{Cd}_6\text{Se}_6$, our predicted QP gaps of 6.61~eV and 5.76~eV are significantly larger than the respective values of 5.90~eV and 5.22~eV reported in an earlier \textit{GW}/BSE study~\cite{noguchi2012}. These variations of the order of 0.5~eV mirror the QP corrections of 4.21~eV and 3.72~eV for $\text{Cd}_3\text{Se}_3$ and $\text{Cd}_6\text{Se}_6$, respectively, compared to the values of 3.73~eV and 3.24~eV obtained by Noguchi \textit{et al.}~\cite{noguchi2012}. These deviations highlight the critical role of numerical convergence in MBPT calculations on low-dimensional systems, which is significantly enhanced by the efficient approach adopted in this work~\cite{Govoni2015}.

\begin{table*}
\centering
\caption{DFT ($E_g^{DFT}$) and QP ($E_g^{GW}$) gaps and their difference corresponding to the QP correction ($\Delta_{\mathrm{QP}} = E_g^{GW} - E_g^{\mathrm{DFT}}$), alongside with the lowest excitation energy computed from the BSE ($E_{\mathrm{BSE}}$) and the associated exciton binding energy: $E_b = E_g^{GW} - E_{\mathrm{BSE}}$. All values are reported in eV.}
\label{tab:gaps_binding}
\begin{tabular*}{0.8\linewidth}{@{\extracolsep{\fill}}lccccc@{}}
\hline \hline
Cluster &
$E_g^{DFT}$ &
$E_g^{GW}$ &
$\Delta_{\mathrm{QP}}$ &
$E_{\mathrm{BSE}}$ &
$E_b$ \\
\hline
Cd$_3$Se$_3$       & 2.39 & 6.61 & 4.21 & 2.54 & 4.07 \\
Cd$_6$Se$_6$       & 2.04 & 5.76 & 3.72 & 2.34 & 3.42 \\
Cd$_{13}$Se$_{13}$ & 2.04 & 5.11 & 3.07 & 2.44 & 2.67 \\
Cd$_{33}$Se$_{33}$ & 1.59 & 4.09 & 2.49 & 2.18 & 1.91 \\
\hline \hline
\end{tabular*}
\end{table*}

\begin{figure}[!ht]
    \centering
    \includegraphics[width=0.45\linewidth]{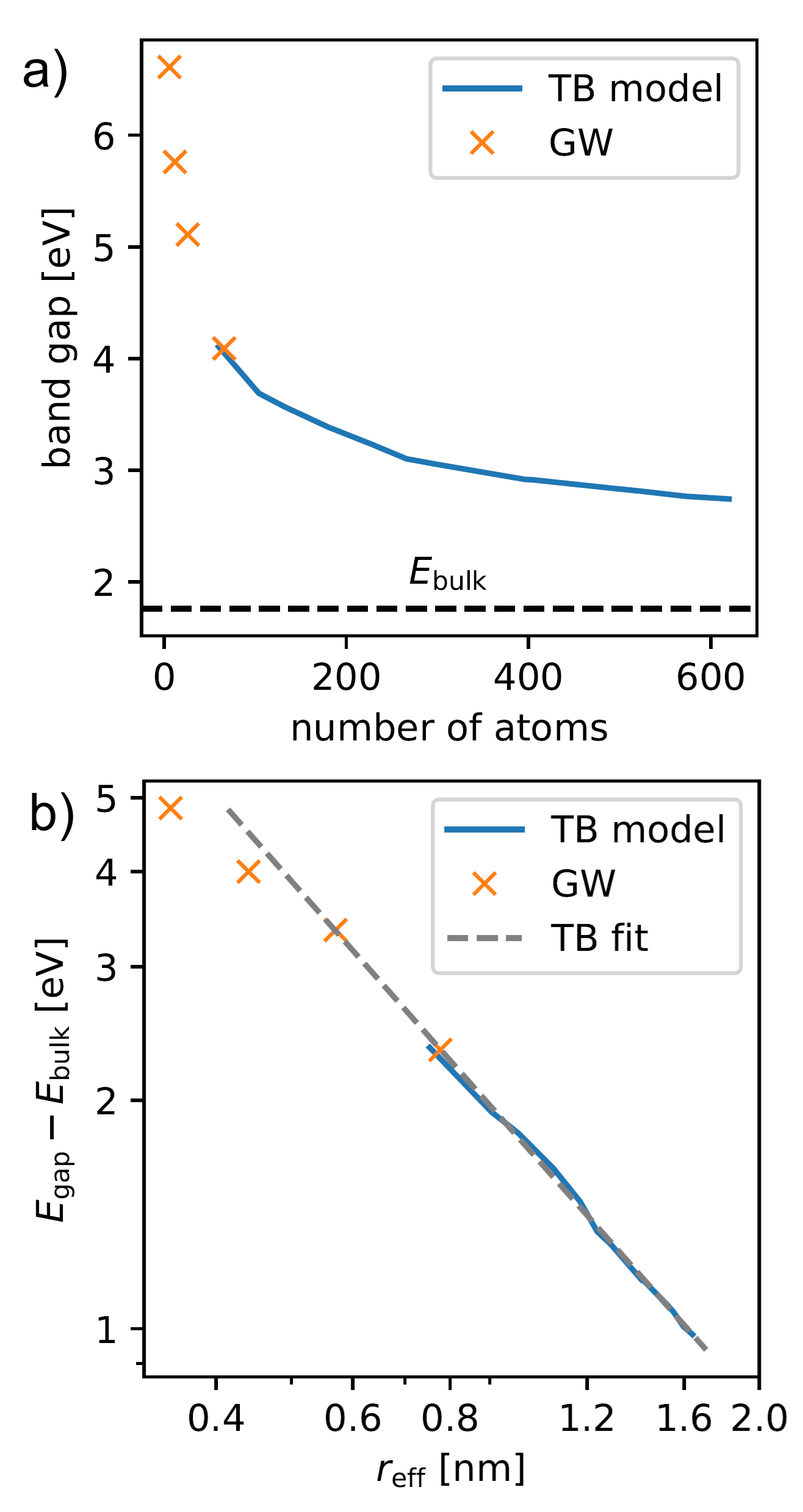}
    \caption{\label{fig:gap-scaling} a) Scaling law of the TB and $G_0W_0$ gap as a function of the number of atoms in the CdSe QDs; the reference band gap in the bulk ($E_{\mathrm{bulk}}=1.76$~eV) is indicated by the dashed horizontal line.
    b) Double logarithmic plot of the confinement energy ($E_{\mathrm{gap}} - E_{\mathrm{bulk}}$) \textit{vs.} the effective radius $r_{\mathrm{eff}}$, with the dashed line indicating a power fit to the TB results.    }
    \label{fig:conv}
\end{figure}

We now compare our $G_0W_0$ results with those obtained with the TB model~\cite{thummler_modeling_2026}, where the bulk band gap is scissor-corrected to match the experimental value of 1.76~eV\cite{ninomiya_optical_1995}. Within this framework, the TB approach captures both quantum confinement effects and the size-dependent QP band-gap scaling. To systematically evaluate this behavior, we define an effective radius ($r_{\mathrm{eff}}$) accounting for the non-spherical geometry of the smaller nanoclusters:
\begin{align}
    r_{\mathrm{eff}} =\sqrt[3]{\frac{3\pi N V}{16}} ,
\end{align}
where $N$ is the number of atoms and $V$ is the unit-cell volume of the bulk wurtzite structure. As shown in Figure~\ref{fig:gap-scaling}a, the TB band gap decreases monotonically with increasing cluster size, gradually approaching the bulk limit. Corrected with a rigid scissors shift of $\epsilon_{\mathrm{shift}}=1.22$~eV, the TB model achieves an excellent quantitative agreement with the $G_0W_0$ prediction for the largest nanocluster ($\mathrm{Cd}_{33}\mathrm{Se}_{33}$). 
 
A double-logarithmic fit of the TB confinement energy (Figure~\ref{fig:gap-scaling}b) yields the power-law relation:
\begin{equation}
E_{\mathrm{gap}} - E_{\mathrm{bulk}} \propto r_{\mathrm{eff}}^{-1.16}.
\label{eq:scaling}
\end{equation}
While being in excellent agreement with previous DFT studies~\cite{Li2005,Dong2024}, Eq.~\eqref{eq:scaling} clearly deviates from the $r^{-2}$ scaling predicted by effective-mass approximations~\cite{Brus1984}. As discussed in Ref.~\citenum{Dong2024}, this non-classical, sub-quadratic relation originates from atomistic quantum confinement effects. For the smallest nanoclusters ($\text{Cd}_3\text{Se}_3$ and $\text{Cd}_6\text{Se}_6$), the TB model slightly overestimates the QP gap compared to the $G_0W_0$ reference. This minor variation stems from the band-structure parametrization from DFT-computed bulk effective masses, which underestimate spatial delocalization under extreme quantum confinement~\cite{Dong2024,kumar2024empirical}.

\begin{figure}[t]
    \centering
    \includegraphics[width=\linewidth]{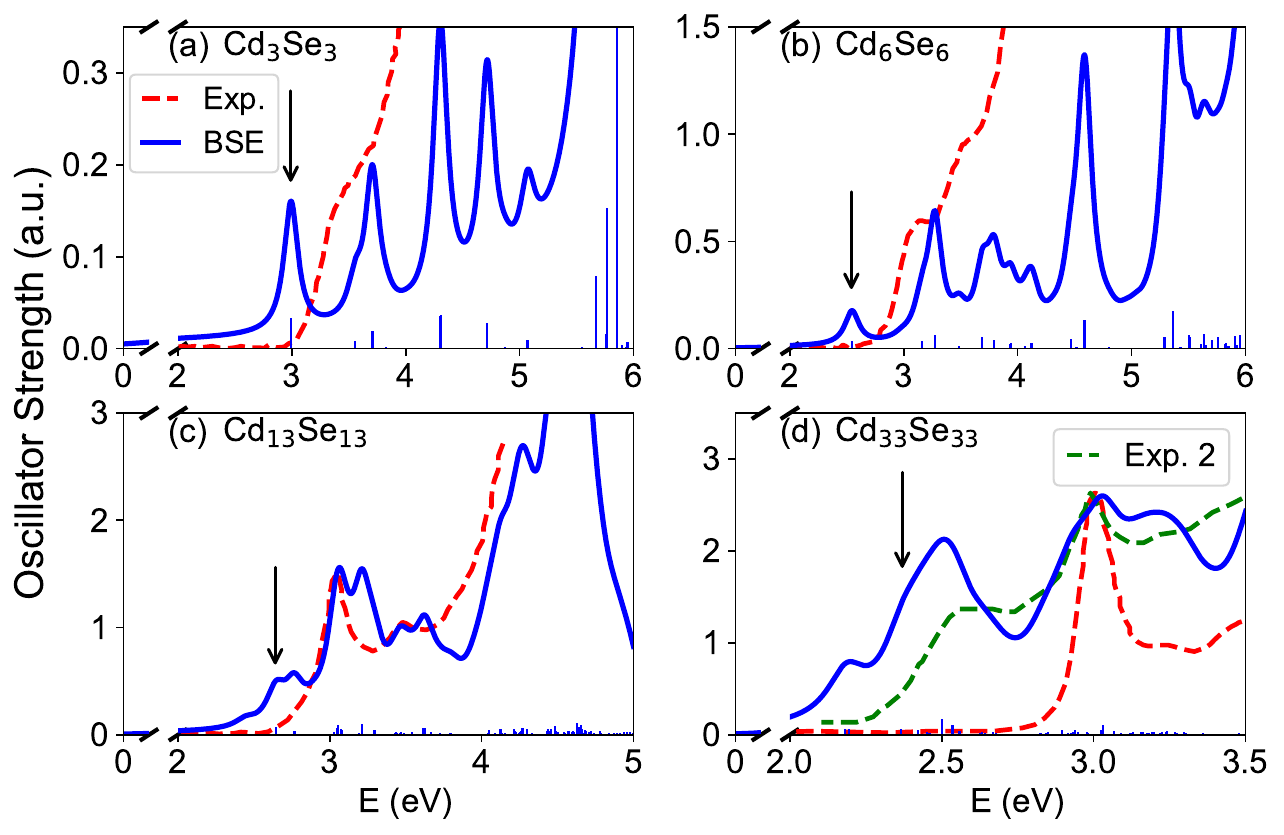}
    \caption{Calculated absorption spectra of (a) Cd$_3$Se$_3$, (b) Cd$_6$Se$_6$, (c) Cd$_{13}$Se$_{13}$, and (d) Cd$_{33}$Se$_{33}$ QDs broadened by a Lorentzian function with full-width at half maximum of 0.15~eV for visualization purposes. The black arrows indicate the energy of the lowest bright excitons in each system. The experimental (Exp.) spectra at 45$^\circ$C for Cd$_3$Se$_3$, Cd$_6$Se$_6$, and Cd$_{13}$Se$_{13}$ are taken from Ref.~\citenum{Jose2006} while for Cd$_{33}$Se$_{33}$, an independent dataset recorded at 80$^\circ$C (Exp. 2)~\cite{kasuya2004}  is added for further comparison. Both experimental datasets are manually extracted from the respective sources and replotted on arbitrary scales.}

   \label{fig:abs}
\end{figure}

To resolve the excitonic resonances and the corresponding transition dipole moment (TDM) strengths, we solved the BSE iteratively in Liouville space using the Davidson algorithm~\cite{Govoni2015}. As a consequence of quantum confinement, the smaller nanoclusters display widely separated excitonic states, whereas the spectrum becomes increasingly dense as the cluster size scales up (Figure~\ref{fig:abs}). In Cd${_3}$Se${_3}$, the lowest excitonic state is found at 2.54~eV, while for Cd$_6$Se$_6$ it occurs at 2.34~eV. These values differ systematically from those reported in Ref.~\citenum{noguchi2012}, where the lowest excitons for these two systems were found at 1.91 and 1.89~eV, respectively. The resulting exciton binding energies, defined as the difference between the $G_0W_0$ gap and the lowest-energy BSE transition ($E_b = E_g^{GW} - E_{\mathrm{BSE}}$) are as large as 4.07~eV and 3.42~eV for Cd$_3$Se$_3$ and Cd$_6$Se$_6$, respectively, in remarkably good agreement with the values of 3.99 and 3.33~eV by Noguchi \textit{et al.}~\cite{noguchi2012}. This accordance demonstrates that the historical 1--2~eV blueshift of \textit{ab initio} MBPT predictions compared to experiments stems most likely from single-particle QP gap convergence rather than from deficiencies in the four-point BSE kernel. The discrepancies with previous results~\cite{noguchi2012} may arise from differences in the starting-point exchange–correlation functionals, plane-wave energy cutoffs, and dielectric screening treatments. In particular, our calculations employ PBE with a converged cutoff of 52 Ry and the PDEP approach, whereas Ref.~\citenum{noguchi2012} uses LDA with a 12.29 Ry cutoff and the generalized plasmon-pole approximation. Although the lower cutoff may contribute to the discrepancies, its precise role remains uncertain without a direct methodological comparison. Further comparison details are provided in the Supporting Information (Section~S5).

For $\text{Cd}_3\text{Se}_3$ and $\text{Cd}_6\text{Se}_6$, the two lowest excitonic states are almost degenerate and optically dark, as a consequence of the near-degeneracy of the HOMO and HOMO-1 orbitals~\cite{Bui2020,Steenbock2023}. However, while earlier predictions placed the first optically active transitions deep in the ultraviolet, namely at 3.75 and 3.57~eV for Cd$_3$Se$_3$ and Cd$_6$Se$_6$, respectively, yielding substantial dark-bright splittings of 1.84 and 1.68~eV~\cite{noguchi2012}, our calculations predict much smaller shifts of 0.45 and 0.20~eV, respectively. 
For $\text{Cd}_{13}\text{Se}_{13}$, the lowest-energy exciton is nearly doubly degenerate (splitting of $\sim$1 meV) and appears at 2.44~eV, matching TDDFT results (2.52~eV)~\cite{gao2014}. Moreover, its oscillator strength remains weak compared to the dominant second exciton at 2.64~eV.
In the larger Cd${_{33}}$Se${_{33}}$ nanocluster, the lowest excitons are found at 2.18 and 2.19~eV, respectively, closely aligning with earlier TDDFT calculations~\cite{Del2011}. Their weak but non-zero oscillator strength marks the onset of structural and electronic evolution toward nanocrystal behavior~\cite{sercel2017}.
This size-dependent evolution of both the QP correction and exciton binding energy highlights a fundamental trend of many-body correlations in nanoconfined systems: the nearly perfect cancellation between electron-electron repulsion, responsible for $\Delta_{\mathrm{QP}}$, and electron–hole attraction ($E_b$)~\cite{Cocchi2015} (Table~\ref{tab:gaps_binding}). In the molecular limit of our smallest cluster ($\text{Cd}_3\text{Se}_3$), the values of $\Delta_{\mathrm{QP}} = 4.21$~eV and $E_b=4.07$~eV almost completely offset each other due to extremely weak dielectric screening. However, as the cluster volume grows, this compensation becomes increasingly asymmetric, with the difference between $\Delta_{\mathrm{QP}}$ and $E_b$ reaching 300~meV and 600~meV in $\text{Cd}_{6}\text{Se}_{6}$ and $\text{Cd}_{13}\text{Se}_{13}$, respectively. In the largest nanocrystal ($\text{Cd}_{33}\text{Se}_{33}$), $E_b$ decreases far more rapidly than $\Delta_{\mathrm{QP}}$ (Table~\ref{tab:gaps_binding}), driven by enhanced exciton delocalization as the cluster of increasing size asymptotically approaches bulk-like characteristics. The breakdown in $\Delta_{\mathrm{QP}}$ \textit{vs.} $E_b$ compensation directly explains the systematic divergence between the BSE optical onsets and the underlying DFT baseline, where the net blue shift relative to the PBE gap grows from 160~meV in $\text{Cd}_3\text{Se}_3$ to 680~meV in $\text{Cd}_{33}\text{Se}_{33}$.

Direct comparison with measurements~\cite{Jose2006,kasuya2004} confirms the overall accuracy of our results (Figure~\ref{fig:abs}). The simulated spectra reproduce well the primary experimental features across all cluster sizes, except for systematic energy shifts ascribed to dissipation effects (solvation, passivation, thermal broadening) not included in DFT+MBPT gas-phase calculations. Across the smaller nanocluster ($\text{Cd}_3\text{Se}_3$, $\text{Cd}_6\text{Se}_6$, and $\text{Cd}_{13}\text{Se}_13$) our BSE calculations resolve weak pre-peaks (indicated in Figure~\ref{fig:abs} by black arrows at 3.0, 2.5, and 2.7~eV, respectively) that precede the steep experimental absorption edge. As substantiated by our IPR analysis (Figure~\ref{fig:stru_level}), these features originate from transitions involving highly localized frontier valence states with weak transition dipole moments. In macroscopic ensembles of disordered or solvated samples, such low-oscillator-strength features are easily obscured by inhomogeneous broadening, causing the experimental absorption threshold to manifest at higher energies where transitions involving deeper, delocalized valence states contribute.

For the largest cluster ($\text{Cd}_{33}\text{Se}_{33}$, Figure~\ref{fig:abs}d), the high density of states gives rise to a smooth absorption onset starting from 2~eV, which culminates in a broad intense resonance at 2.5~eV. Interestingly, this feature aligns with an experimental spectrum recorded at 80$^{\circ}$C~\cite{kasuya2004} (green dashed curve), whereas ambient-condition measurements~\cite{Jose2006} exhibit a sharper onset near 2.9~eV (red dashed curve). This agreement suggests that elevated temperatures induce structural fluctuations and phonon-assisted transitions that thermally activate weak near-gap excitations, reproducing the absorption profile predicted by our zero-temperature BSE calculations. 
Ligands and solvation effects further contribute to the measured spectra~\cite{Kalyuzhny2005,Azpiroz2013,SUN2016,fenoll2023,Fischer2012}, by modifying the character of frontier orbitals and shifting the absorption onset by 0.2-0.4 eV, as demonstrated in previous theoretical work~\cite{Azpiroz2013,SUN2016,fenoll2023,Fischer2012}, even though ligand-derived states typically lie deeper in the occupied or higher in the unoccupied electronic spectrum \cite{Fischer2012}. 
\begin{table}[h]
\centering
\caption{Dominant ($> 10\%$) single-particle contributions to the lowest 5 excitons in four CdSe clusters with the corresponding percent weights reported in parentheses. H and L stand for HOMO and LUMO, respectively.}
\begin{tabular}{c c c p{7.5cm}}
\hline
Cluster & Exciton (eV) & Oscillator strength & Dominant transitions (weight \%) \\
\hline
             & 2.54 & $4.08\cdot 10^{-7}$ & H-1$\rightarrow$L (84.7) \\
             & 2.54 & $3.70\cdot 10^{-7}$ & H$\rightarrow$L (84.5) \\
Cd$_3$Se$_3$ & 2.99 & $3.33\cdot 10^{-2}$ & H-2$\rightarrow$L (82.8) \\
             & 3.09 & $1.16\cdot 10^{-5}$ & H$\rightarrow$L+1 (79.9), H-1$\rightarrow$L+1 (15.6) \\
             & 3.09 & $5.65\cdot 10^{-6}$ & H-1$\rightarrow$L+1 (80.1), H$\rightarrow$L+1 (15.6) \\
\hline 
             & 2.34 & $4.28\cdot 10^{-6}$ & H$\rightarrow$L (62.8), H-1$\rightarrow$L (32.6) \\
             & 2.34 & $4.05\cdot 10^{-6}$ & H-1$\rightarrow$L (62.9), H$\rightarrow$L (32.6) \\
Cd$_6$Se$_6$ & 2.54 & $3.65\cdot 10^{-2}$ & H-2$\rightarrow$L (92.0) \\
             & 2.96 & $1.59\cdot 10^{-5}$ & H-1$\rightarrow$L (47.3), H$\rightarrow$L (36.8) \\
             & 2.96 & $5.45\cdot 10^{-6}$ & H$\rightarrow$L (46.8), H-1$\rightarrow$L (36.9) \\
\hline
             & 2.44 & $7.66\cdot 10^{-3}$ & H-1$\rightarrow$L (87.1) \\
             & 2.44 & $7.68\cdot 10^{-3}$ & H$\rightarrow$L (87.0) \\
Cd$_{13}$Se$_{13}$ & 2.64 & $7.20\cdot 10^{-2}$ & H-2$\rightarrow$L (84.5) \\
             & 2.68 & $1.18\cdot 10^{-4}$ & H-3$\rightarrow$L (87.3) \\
             & 2.76 & $2.12\cdot 10^{-2}$ & H-2$\rightarrow$L (85.0) \\

\hline
             & 2.18 & $2.35\cdot 10^{-2}$ & H$\rightarrow$L (92.5) \\
             & 2.19 & $3.50\cdot 10^{-2}$ & H-1$\rightarrow$L (92.4) \\
Cd$_{33}\text{Se}_{33}$ & 2.34 & $2.61\cdot 10^{-2}$ & H-4$\rightarrow$L (52.8), H-3$\rightarrow$L (18.9) \\
             & 2.37 & $4.38\cdot 10^{-1}$ & H-3$\rightarrow$L (63.8), H-4$\rightarrow$L (15.1) \\
             & 2.37 & $6.47\cdot 10^{-1}$ & H-2$\rightarrow$L (73.2), H-5$\rightarrow$L (10.2) \\
\hline

\end{tabular}
\label{tab:weights}
\end{table}

To further uncover the microscopic mechanism behind the optical suppression of these pre-peak features, we project the BSE excitonic amplitudes onto the underlying Kohn–Sham single-particle basis (Table~\ref{tab:weights}). Across all cluster sizes, the lowest excitonic transitions, ranging from 2.54~eV in $\text{Cd}_3\text{Se}_3$ to 2.18–2.19~eV in $\text{Cd}_{33}\text{Se}_{33}$, are dominated almost exclusively by transitions between the highest occupied ($\text{HOMO}/\text{HOMO}-1$) and lowest unoccupied state ($\text{LUMO}$) manifolds.
This quenching stems directly from the spatial wave-function mismatch resolved in our IPR analysis (Figure~\ref{fig:stru_level}). Across the entire size series, valence states close to the frontier consistently exhibit high IPR values, indicating pronounced localization~\cite{goldzak2021,mcisaac2023,steenbock2024}, whereas near-gap conduction states are highly delocalized within the core. This spatial mismatch dramatically quenches the TDM, rendering these optical transitions at the onset virtually dark in experiments, where optical absorption is only activated at higher energies by deeper, more delocalized valence states that restore spatial overlap.


%

Structural symmetry governs fine structure and optical selection rules of these threshold excitons. The high symmetry of the three smallest clusters ($\text{Cd}_3\text{Se}_3$, $\text{Cd}_6\text{Se}_6$, and $\text{Cd}_{13}\text{Se}_{13}$) enforces a strict degeneracy within their frontier valence states, giving rise to perfectly degenerate excitonic pairs (Table~\ref{tab:weights}). In $\text{Cd}_{33}\text{Se}_{33}$, the emergence of a bulk-like wurtzite core breaks this symmetry, lifting the HOMO degeneracy into two distinct levels at 2.18~eV and 2.19~eV and allowing these asymmetric transitions to pick up a small, non-zero oscillator strength that manifests as a faint pre-onset in high-resolution measurements~\cite{kasuya2004,botti2007}. As the cluster size expands further, the rapidly growing density of states and increased excitonic configuration mixing drives a crossover from discrete QD-like transitions to broad nanocrystal absorption profiles.

\begin{figure}
    \centering
    \includegraphics[width=0.5\linewidth]{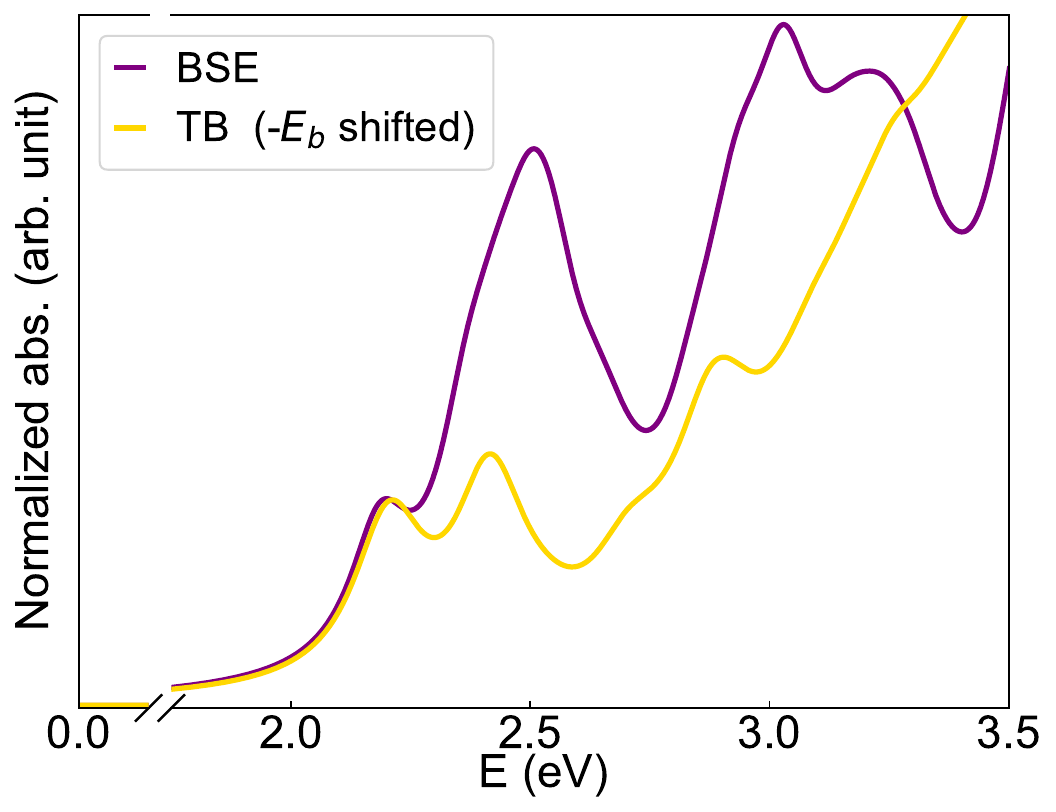}
   \caption{Optical absorption spectra calculated from $GW$/BSE (purple) ($\mathrm{Cd}_{33}\mathrm{Se}_{33}$) and the TB model ($\mathrm{Cd}_{30}\mathrm{Se}_{30}$) rigidly shifted by the lowest exciton binding energy ($E_b$) obtained from our BSE results. A Lorentzian broadening of 0.15~eV is applied to both spectra.}

    \label{fig:TB_GW}
\end{figure}

To evaluate the transferability of our scalable atomistic framework, we compare the optical response obtained from our TB model for the $\mathrm{Cd}_{30}\mathrm{Se}_{30}$ nanocluster~\cite{thummler_modeling_2026} against the $G_0W_0$/BSE benchmark $\mathrm{Cd}_{33}\mathrm{Se}_{33}$ (Figure~\ref{fig:TB_GW}). To compensate for the missing electron--hole interactions in our TB model, we rigidly shift the corresponding spectrum by the lowest-exciton binding energy ($E_b$) obtained from BSE (Table~\ref{tab:gaps_binding}), thereby aligning the absorption onsets. 

The excitation energies in the two spectra agree fairly well even beyond the first aligned peak, with differences up to 200~meV (Figure~\ref{fig:TB_GW}). This mismatch is not surprising, considering that the exciton binding energy is a state-dependent quantity~\cite{Cocchi2015} and the applied rigid shift does not reproduce variations directly associated with electron-hole correlations. It is worth recalling these effects are not included in our TB model, particularly the mixing of single-particle transitions responsible for the redistribution of the oscillator strength above the onset. On the other hand, the lowest bright excitons in GW/BSE are mainly dominated by individual transitions, explaining the good agreement for the first peaks. Further details are reported in the Supporting Information (Section S3). The inclusion of only two unoccupied bands in the transition space (Figure~S3) further contributes to the systematically lower oscillator strength in the TB-model spectra compared to the BSE results. This choice depletes the number of available transitions, consequently decreasing the overall spectral weight. Altogether, this analysis confirms the ability of the TB model to capture the general low-energy features of a fully \textit{ab initio} $GW$/BSE spectrum. A promising route toward achieving MBPT-level accuracy would be to combine the TB electronic structure with a screened configuration-interaction (CI)\cite{Bui2020,Steenbock2023} framework, employing an analytical dielectric screening function to describe the screened electron–hole Coulomb interaction. This approach could enable accurate calculations of the optical response of larger nanocrystals that remain computationally inaccessible to conventional MBPT methods.
\section{CONCLUSION}
In summary, we have presented a comprehensive first-principles many-body perturbation theory study ($GW$/BSE) of stoichiometric Cd$_n$Se$_n$ nanoclusters ($n=3$, 6, 13, and 33) and leveraged it to validate a scalable atomistic TB model. Supported by very good agreement with experimental references and earlier MBPT studies, we clarified remaining discrepancies in the literature. By explicitly evaluating exciton binding energies, we demonstrate that historical 1--2~eV blueshifts reported in earlier MBPT studies stem from underconverged $GW$ calculations rather than deficiencies of electron–hole kernels. We found that the systematic cancellation between QP corrections and exciton binding energies due to quantum confinement breaks down as cluster volume increases, due to the onset of bulk-like dielectric screening that attenuates $E_b$ faster than $\Delta_{\mathrm{QP}}$. This trend explains why mean-field DFT gaps deceptively mirror optical thresholds in small clusters but systematically diverge by several hundreds of meV in larger nanostructures. IPR analysis on the underlying electronic structure explains the origin of the optical suppression of pre-peaks: a severe spatial wave-function mismatch between localized frontier valence orbitals and delocalized core conduction states quenches the transition dipole moments in the smallest clusters. As the volume expands, structural symmetry breaking and a dense orbital manifold restore spatial overlap, driving an excitonic transition from discrete molecular transitions to broad, bulk-like optical profiles. 

Finally, we demonstrate that a scissor-corrected, DFT-parameterized TB model accurately captures both the non-trivial confinement scaling and the low-energy optical response, drastically outperforming conventional effective-mass approximations. By identifying the exact physical limits where independent-particle tight-binding breaks down due to state-dependent exciton binding energies, this work establishes a clear multiscale roadmap: combining scalable tight-binding backbones with effective two-particle kernels enables predictive many-body simulations of linear and nonlinear optical phenomena in semiconductor nanostructures containing thousands of atoms.
Building upon these findings, future work will explore atom-centered orbital bases, environmentally adaptive tight-binding parameters, and localized surface potentials to enable a more detailed and accurate description of surface chemistry and ligand-nanocrystal interactions, ultimately paving the way for fully atomistic multiscale simulations of complex, realistic morphologies.

\begin{acknowledgement}
This work was funded by the German Research Foundation, Project number 398816777 (CRC 1375, sub-projects A1 and A8). Computational resources were provided by the German National High-Performance Computing Alliance (NHR), project FUSE-TOP (thp00002). 

\end{acknowledgement}

\begin{suppinfo}
The following files are available free of charge. The SI includes: 
\begin{itemize}
\item SI.pdf : This file contains convergence tests, theoretical foundation of the tight-binding model along with size-dependent absorption spectra with the model, comparison of GW/BSE and TDDFT absorption spectra and   a discussion of potential source for discrepancies in quasiparticle results.  
\end{itemize}
\end{suppinfo}
\vspace{0.5 cm}

\noindent\textbf{Data Availability Statement}\\
Input and output files of the ab initio calculations performed in this work are available free of charge in Zenodo \href{https://doi.org/10.5281/zenodo.21773808}{DOI: 10.5281/zenodo.21773808} [record: 21773808]. 
\vspace{0.5 cm}

\noindent\textbf{Notes}\\
The authors declare no competing financial interest.

\bibliography{main}

\end{document}